\documentclass[twocolumn,showpacs,preprintnumbers,amsmath,amssymb]{revtex4}
\usepackage{graphicx}
\usepackage{dcolumn}
\usepackage{bm}

\begin{document}
\preprint{APS/123-QED}
\title{New excitations in bcc $^{4}$He - an inelastic neutron
scattering study}

\author{O. Pelleg$^{1}$}\email{poshri@tx.technion.ac.il}\homepage{http:\\physics.technion.ac.il\~poshri}
\author{J. Bossy$^{2}$} \author{E. Farhi$^{3}$} \author{M. Shay$^{1}$} \author{V. Sorkin$^{1}$}
\author{E. Polturak$^{1}$} \email{emilp@physics.technion.ac.il}
\homepage{http://physics.technion.ac.il/~emilp/}
\address{(1)Physics Department, Technion - IIT, Haifa, Israel 3200, \\
(2) CNRS-CRTBT, BP166, 38042 Grenoble Cedex 9,France, \\
(3) Institut Laue Langevin, BP 156, 38042 Grenoble Cedex
9,France.}
\date{\today}

\pacs{67.80.Cx, 67.80.-s, 63.20.Dj, 63.20. Pw}


\begin{abstract}
We report neutron scattering measurements on bcc solid $^{4}$%
He. We studied the phonon branches and the recently discovered
''optic-like'' branch along the main crystalline directions. In
addition, we discovered another, dispersionless "optic-like''
branch  at an energy around 1 meV ($\sim$~11K). The properties of
the two "optic-like" branches seem different. Since one expects
only 3 acoustic phonon branches in a monoatomic cubic crystal,
these new branches must represent different type of excitations.
One possible interpretation involves localized excitations unique
to a quantum solid.
\end{abstract}

\maketitle

Elementary excitations of quantum solids show unique properties
arising from the large zero point vibration of the atoms. The
interatomic potential is highly anharmonic, and the atoms feel
strong short range correlations, due to the repulsive part of the
potential\cite {koehler,varma,glydebook}. The "self consistent
phonon" theory developed to treat this problem yielded phonon
dispersion curves in a reasonable agreement with experimental data
available at the time\cite{minkiewicz,osgood}. Recently, an
additional, ''optic-like'' excitation branch was
discovered\cite{tuvy} along the [110] direction of bcc $^{4}$He.
This is puzzling since on general grounds, only \textit{acoustic}
phonon branches should exist in a mono-atomic cubic crystal.
Several interpretations of the new excitation were
discussed\cite{tuvy} in terms of multiphonon
effects\cite{glydebook}, or localized excitations unique to a
quantum solid. The latter include an isotropic vacancy
band\cite{hetherington,guyer2,andreev} or anisotropic local modes
in the [110] direction associated with correlated zero point
motion\cite{gov}. An analysis of the data showed that multiphonon
effects were not the source of this new branch\cite{tuvy}. In
addition, each of the models involving localized excitations was
consistent only with some facet of the data\cite{tuvy}. Localized
excitations in solid He are of particular interest in view of the
recent reports of the "supersolid" phase\cite{chan}. Our
motivation to do additional experiments was twofold: first, the
new ''optic-like'' branch was measured only along the [110]
crystalline direction. The anisotropy of this new branch can be
determined only if it is investigated in all directions. Second,
since the new branch was largely unexpected, perhaps other such
features may be found.

\begin{figure*}[t!]
\includegraphics[width=13cm]{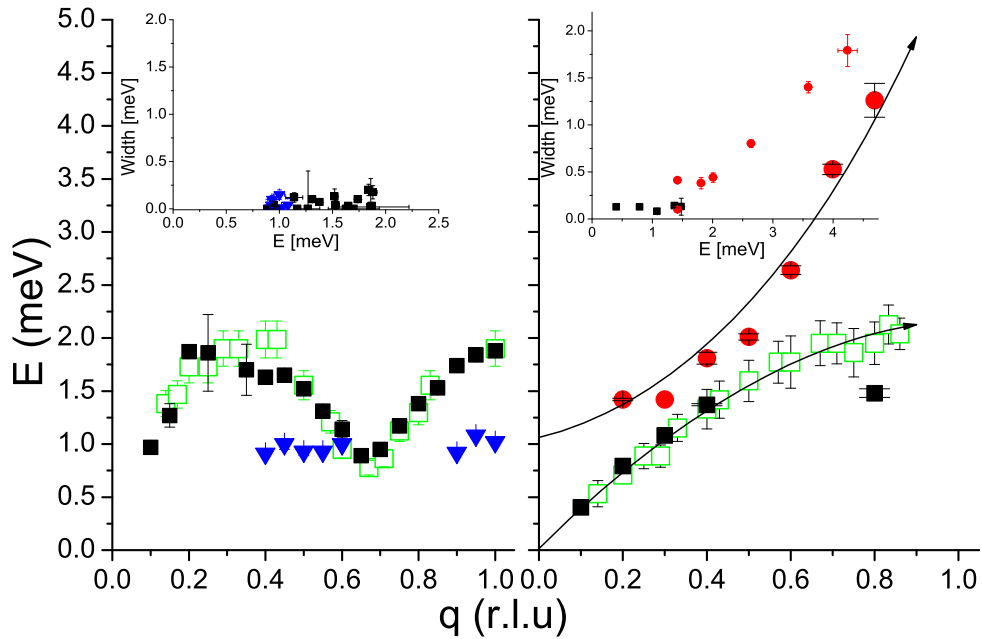}
\caption{\label{fig1}(color online) Dispersions curves along the
[111] direction. Left panel: longitudinal polarization. Right
panel: transverse polarization. In both panels, solid black
squares are data points for the phonon branches, and the open
green squares are results of PIMC simulations of the dispersion
curves\cite{slava}. The inverted blue triangles in the left panel
are data points for the LOB. In the right panel, the red circles
are data points for the HOB. The solid lines are a guide to the
eye. The inset shows the energy linewidth of the features seen in
the inelastic spectra. Note that broadening occurs only for the
polarization where the HOB is observed(right panel).}
\end{figure*}

In that spirit, we performed inelastic neutron scattering
experiments on bcc $^{4}$He using the IN-12 and IN-14 triple axis
spectrometers at the ILL. Single crystals of high purity $^{4}$He
(less than 0.1 ppm $^{3}$He) were grown and oriented in the beam.
The growth orientation of these crystals was a priori random. In
most cases, the limitations of orienting the crystals by tilting
the cryostat allowed us to measure only along the [110] direction.
In order to access other crystallographic directions, we added a
home built cold goniometer to the sample stage. Two versions of
the apparatus were built, the first one allowing an additional
$\pm$12${^\circ}$ tilt of the cell in an arbitrary direction. This
apparatus was used for inelastic scattering along the [110]
direction carried out on the IN-14 spectrometer. The second
version, allowing an additional $\pm$45${^\circ}$ tilt along one
axis, was used for measurements along [100] and [111] done in a
separate experiment on the IN-12 spectrometer. Crystals of low
density $^{4}$He (21 cm$^3$ molar volume), several cm$^3$ in size
were grown from the superfluid at T=1.640 K, where the temperature
width of the bcc phase is maximal($\sim$ 50 mK). In order to get a
high quality single crystal, the solid was further annealed
overnight on the melting curve. The final crystal was composed of
two large grains misaligned by 20', with volumes of about 2 and 4
cm$^3$. The FWHM of the rocking curve was about 30' and 40' for
the two grains. Scattering experiments were conducted at the same
temperature. In addition to the crystal, the cell contained a
small amount of superfluid helium (less then 1$\%$ of the volume)
in order to reduce any temperature gradients across the sample. In
particular, we were able to get very high resolution data along
[001] and [111] from the \textit{same} crystal, which could be
oriented to access either the \{002\} scattering plane or the
\{1$\overline{1}$2\}. The energy of the incident beam could be
varied in the range of 2.3 to 14 meV. Constant-Q scans were done
at a fixed momentum of the scattered neutrons (k$_{F}$). In the
high resolution scans we used a cooled Be filter to remove
$\lambda /2$ contamination. Focusing techniques were used to
enhance the scattering intensity. The highest instrumental
resolution was 0.1~meV (FWHM) with k$_{F}$=1.2 $\rm\AA^{-1}$. In
order to ascertain that the new features described in the
following are not spurious, we carried out extensive background
measurements by repeating the inelastic scans in the various
directions once with an empty cell, and once with the cell full of
liquid. We found that the new features disappeared once the
crystal was molten, hence they're not connected with scattering
from the liquid or from the walls of the cell. Bragg scattering
measurements of the crystals failed to show any traces of the hcp
phase. We conclude therefore that the new features are a property
of the bcc solid.

Our measurements of the phonons fill some gaps in the available
data for the phonon dispersion
curves\cite{minkiewicz,osgood,tuvyJLTP}. These results will be
described in detail elsewhere\cite{to be published}. In this
paper, we focus on the new optic-like branches. In order to assign
the various peaks seen in the neutron scattering data, we
performed Path Integral Monte Carlo (PIMC) simulations\cite{slava}
of bcc $^4$He to determine the contribution of single phonons to
the dynamic structure factor S(q,$\omega$). Figure ~\ref{fig1}
shows the dispersion relations of the L and T phonons along [111],
along with the two "optic-like" branches observed in these
experiment. The left panel shows the L branch and another,
dispersionless branch at an energy around 1 meV($\sim$~11K). We
label this new branch as the "lower optic-like branch"(LOB).
Regarding the phonons, the data for the L branch is in excellent
agreement with the simulations. The measured energy linewidth of
the phonon peaks in the spectra (see inset) is small, limited by
the experimental resolution. Turing to the right panel of Figure
~\ref{fig1}, again two branches are seen, the T[111] phonon branch
and another, "optic-like" branch which we identify with the branch
observed previously\cite{tuvy} along [110]. We label this branch
as the "higher optic-like branch"(HOB). It is seen that the data
for the T branch agrees with the simulations only for energies
less than the minimum energy of the HOB branch. Similarly, the
inset shows that the linewidth increases quite strongly above this
minimum energy. Quite obviously, one cannot assign the points on
the dispersion curve with a large linewidth to single phonons. The
increase of the linewidth was observed also for the L[100] branch
where the HOB excitation was also seen.

\begin{figure}[h!]
\includegraphics[width=7cm]{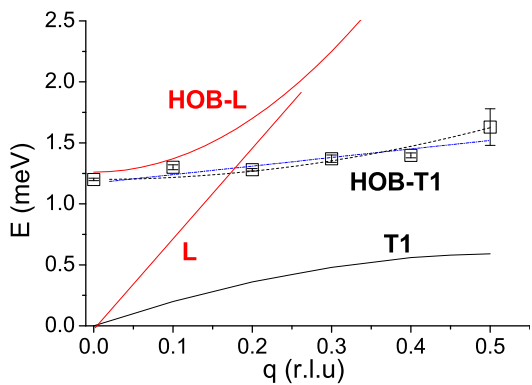}
\caption{\label{fig2}(color online) Dispersion curves of the
Higher Optical Branch (HOB) along [110] with two polarizations.
The lines labelled HOB-L and L are smoothed values of the HOB and
the L(110) phonon branches from Reference \cite{tuvy}. HOB-T1
denotes data points from this work for the HOB with T1
polarization, while the line labelled T1 represents smoothed
values of the T1(110) phonon branch. The dashed black line is a
linear fit to the HOB-T1 data, and the dash-dot blue line is a
 quadratic fit.}
\end{figure}

First, we discuss the HOB ''optic-like'' excitation branch,
previously observed only along the [110] direction\cite{tuvy}. In
the present work, this branch was observed in scans measuring the
T(111) and L(100) phonons, while in the scans for L(111) and
T(100) it was absent. Hence, this excitation is anisotropic. The
dispersion of this branch, when plotted together with the usual
phonons is suggestive of mode coupling. This is evident for
example in the right panel of Figure~\ref{fig1} showing the T[111]
branch, as well as in earlier data\cite{tuvy} along [110].
Additional support for this suggestion comes from the results
plotted in Figure~\ref{fig2} showing the polarization dependence
of the HOB along [110]. It is seen that in scans with the T1
polarization, this branch (labelled HOB-T1) shows little
dispersion, while in scans with a longitudinal polarization
(labelled HOB-L) the dispersion is significant. Since the maximal
energy of the T1[110] branch is 0.6 meV ($\sim$~7K), while the
minimal energy of the HOB is 1.23 meV ($\sim$~14K), the T1 and HOB
branches do not cross. Hence, mode coupling with the T1 branch
should be weak. On the other hand, the HOB and the L(110) branch
do cross and so the coupling with the L(110) should be stronger.
Both the broadening shown in the right panel of Figure~\ref{fig1}
and the data in Figure~\ref{fig2} are consistent with this idea.

In order to determine the intrinsic dispersion of this branch, we
attempted to simultaneously fit the dispersion relations of the
HOB excitation and the phonon branches using a mode coupling
approach. We took the coupling of the branches to be in the form
$\Delta/((\omega_{HOB}^2-\omega_{ph}^2) + iD)$, where $\Delta$ is
the coupling constant, $\omega_{HOB}$ and $\omega_{ph}$ denote the
energies of the HOB and phonon branches respectively, and D is
connected to the damping which prevents a singularity at mode
crossing. This particular form is not based on a specific model of
this excitation, hence it is useful only to gauge the relative
strength of the coupling between the various branches. After
trying several similar forms\cite{wood,pippard}, two conclusions
can be drawn; first, it was possible to fit the coupled branches
only by assuming that the HOB has a finite dispersion. Second, the
value of the coupling constant $\Delta$ is different for different
phonon branches. For the [110] direction, it is about twice as big
as for the other directions. The ratio $\Delta/D$ however, is
approximately constant for all the branches. Consequently, we
believe that the intrinsic dispersion of the HOB is that observed
with the T1(110) polarization (Fig.~\ref{fig2}). The dispersion is
weak, and within our resolution can be fitted either to a linear
or quadratic dependence on q. The linear fit, E(q)=$\epsilon
_{0}+v_G q$, yields a group velocity $v_G \sim$~68~m/sec. The
quadratic fit, E(q)=$\epsilon _{0}+\hbar ^{2}q^{2}/2m^{\ast }$,
appears marginally better, with $\epsilon _{0}$~=1.20~meV and
$m^{\ast }\simeq$~0.7~m$_{4}$. The total bandwidth is about 0.4
meV ($\sim$~4.6K). To conclude, the HOB branch is anisotropic, in
the sense that is observed only in certain directions, and seems
to couple more strongly to phonons along [110]. This branch has a
finite intrinsic dispersion.

\begin{figure}[h!]
\includegraphics[width=6cm]{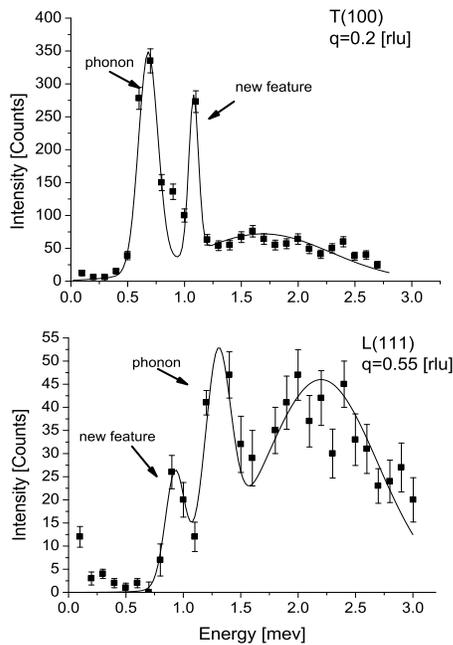}
\caption{\label{fig3} Neutron groups in the [100] and [111]
directions showing the new low energy "optic like" (LOB)
excitation. The squares are data points and the solid line is a
gaussian fit. The "new feature" marks the LOB. The instrumental
resolution in the upper panel is higher.}
\end{figure}

We now discuss the second new feature, namely the LOB (see left
panel,Figure~\ref{fig1}). This excitation branch appears
dispersionless within our resolution, with an energy of
0.95$\pm$0.1 meV [$\sim$~11K]. Typical scans showing the LOB are
plotted in figure~\ref{fig3}. This new excitation was seen in
scans measuring the T(100), L(100) and the L[111] phonons while
being absent in scans along the [110] direction, even at the
highest resolution\cite{tuvy}. Hence, it is also anisotropic in
the same sense as the HOB branch. In this context, we also tried
to examine the possibility that this mode is excited only when the
incident beam is along some specific direction. We found that in
the \{002\} scattering plane, the excitation was observed when the
incident beam was around the (110) direction, while in the
\{1$\overline{1}$2\} plane the incident beam was around the (021)
direction. These directions are not equivalent, so the excitation
of the mode does not seem to be linked to the direction of the
incident neutrons. Another remark is that near the minimum of the
dispersion curve of L[111] (q=2/3 r.l.u.), there is an inherent
mixing of the L[111] phonons with phonons originating in the
\{112\} scattering planes. Hence, these planes may contribute to
the intensity of the L[111] phonons and perhaps add another
feature to the scans. However, the LOB is seen also in the [100]
direction, where there is no phonon mixing. Additional discussion
of these points will be presented elsewhere\cite{to be published}.
The energy linewidth of this new feature is small, limited by the
instrumental resolution. The intensity of the LOB decreases with
reduced resolution, and at low resolution it was not observed.
This may perhaps explain why it was not seen in the
past\cite{minkiewicz}. It may also be the reason why the LOB was
not seen in the scans measuring the T[111] branch, where we were
not able to work with a high resolution. The fact that the LOB is
dispersionless and its linewidth remains small even when it
crosses a phonon branch implies that there is no interaction
between the LOB and phonons. In that respect, the LOB and the HOB
are different types of excitations.

The absence of dispersion of the LOB suggests that this excitation
is localized. Localized excitations can be point defects or more
complex entities, e.g. ''local modes''\cite{sievers}. In usual
materials, phonon energies are in the meV range while those of
point defects are in the eV range. Consequently, in usual
materials point defects cannot be observed in scans measuring
phonon branches. In solid He, these two energy scales are very
similar, so in principle point defects could be excited by cold
neutrons. The energy of the LOB is indeed similar to the typical
energy of a point defect in bcc $^4$He measured by different
techniques\cite{simmons,goodkind}. A neutron incident on the solid
can create a vacancy by knocking a single atom away from its
lattice site. However, single particle excitations have a large
energy width because the final state of the atom is in the
continuum\cite{minkiewicz}. Hence, this possibility is not
consistent with the small linewidth of the LOB. The same argument
applies to creation of vacancy-interstitial pairs (Frenkel pairs).
The formation energy of a Frenkel pair depends on the
interstitial-vacancy distance, which can take different values.
Thus, one expects a broad feature in the scattering intensity vs.
energy rather than the narrow peak which is observed
(Fig.~\ref{fig3}). In addition, recent simulations of point
defects in bcc He indicate that a vacancy branch has a
considerable dispersion\cite{galli}. This result supports our
claim that the dispersionless LOB branch is probably not
associated with creation of vacancies. Another possibility is that
the neutrons excite some resonant mode\cite{sievers, dederichs},
namely internal vibrations of split interstitials or vacancies
already present in the crystal. Formation energies and resonant
mode energies of point defects should be similar, and in addition,
the energy width of resonant modes is small. Some qualitative
evidence of the presence of such excitations was observed in our
simulations of crystals containing a large number of
interstitials\cite{to be published}. Hence, although we did not
find any analytical calculations of these modes for solid He, this
possibility is of interest. Finally, a feature similar to the LOB
may have been observed in neutron scattering from the hcp solid
phase at a temperature of 100 mK\cite{bossy}. Resonant modes, if
they exist, should be observed in both hcp and bcc He. Further
investigation of these features in both solid phases may help to
understand their origin.

In conclusion, we investigated the recently discovered
"optic-like" excitation branch(HOB) and found it to be anisotropic
and weakly dispersive. In addition, another new excitation branch
was discovered (LOB). This branch is also anisotropic,
dispersionless, and with a very small linewidth. In contrast with
the HOB, the LOB does not couple to phonons. One possible
interpretation is that these branches are associated with point
defects or excitations thereof. However, none of the existing
models is detailed enough to allow meaningful comparison with the
data.

We thank S. Hoida (Technion), S. Raymond, A. Stunault, F. Thomas,
S. Pujol and J. Previtali (ILL) for their invaluable contribution
to this experiment. We have benefitted from several discussions
with N. Gov.  This work was supported in part by The Israel
Science Foundation, and by the Technion Fund for Promotion of
Research.

\bigskip

%
%


\bigskip

\begin{references}


\bibitem{koehler}  T.R. Koehler, In {\it Dynamical Properties of Solids, }%
Vol II, (G.K. Horton and A.A. Maradudin, Ed.), Chap. 1,
North-Holland, Amsterdam (1975).
\bibitem{varma}  C.M. Varma and N.R. Werthamer, In {\it The Physics of
Liquid and Solid Helium}, Vol. 1, (K.H. Benneman and J.B.
Ketterson, Eds.), p. 503, Wiley, New York (1976).
\bibitem{glydebook}  H.R. Glyde, {\it Excitations in Liquid and Solid Helium}%
, Clarendon Press, Oxford (1994).
\bibitem{minkiewicz}  V. J. Minkiewicz, T. A. Kitchens, G. Shirane, and E.
B. Osgood, Phys. Rev. {\bf A8}, 1513 (1973).
\bibitem{osgood}  E.B. Osgood, V.J. Minkiewicz, T.A. Kitchens and G.
Shirane, Phys. Rev. {\bf A5}, 1537, (1972).
\bibitem{tuvy} T. Markovich, E. Polturak, J. Bossy and E. Farhi, \textit{Phys. Rev. Lett.}
\textbf{88}, 195301 (2002).

\bibitem{hetherington}  J.H. Hetherington, Phys. Rev. {\bf 176}, 231 (1968).
\bibitem{guyer2}  R.H. Guyer, Jour. Low Temp. Phys. {\bf 8}, 427 (1972).
\bibitem{andreev}  A.F. Andreev, In {\it Progress in Low Temperature Physics}%
, Vol VIII (D.F. Brewer, Ed.), Chap. 2, North-Holland (1982).
\bibitem{gov}  N. Gov and E. Polturak, Phys. Rev. {\bf B60}, 1019 (1999).
\bibitem{chan} E. Kim and M.H.W. Chan, \textit{Science} \textbf{305},1941 (2004).
\bibitem{tuvyJLTP} T. Markovich, E. Polturak, S.G. Lipson, J.Bossy, E. Farhi, M.J. Harris, and M.J. Bull
\textit{Jour. Low Temp. Phys.} \textbf{129}, 65(2002)
\bibitem{to be published} O. Pelleg, J. Bossy, E. Farhi, M. Shay, V. Sorkin and E.
Polturak, to be published.
\bibitem{slava} V.Sorkin, E. Polturak and Joan Adler, \textit{Phys. Rev. B} \textbf{71}, 214304 (2005)
\bibitem{wood} R.F. Wood and M. Mostoller, Phys. Rev. Lett. {\bf 35}, 45
(1975).
\bibitem{pippard} A. B. Pippard, "The Physics of vibrations",
Cambridge University Press (London), 1978.
\bibitem{sievers}  M. Sato, A. J. Sievers, Nature {\bf 432}, 486 (2004), A. J. Sievers,
and S. Takeno, Phys. Rev. Lett. {\bf 61}, 970 (1988).
\bibitem{simmons}  B. A. Fraass, P. R. Granfors, and R. O. Simmons , Phys. Rev. {\bf 39}, 124 (1989),
\bibitem{goodkind} C. A. Burns and J. M. Goodkind, \textit{J. Low Temp.
Phys.} \textbf{95}, 695(1994).
\bibitem{galli} D. E. Galli and L.
Reatto, \textit{J. Low Temp. Phys.} \textbf{134}, 121(2004).
\bibitem{dederichs} P.H. Dederichs, C. Lehmann, A. Scholtz,
\textit{Phys. Rev. Lett.} \textbf{31}, 1130 (1973)
\bibitem{bossy} J. Bossy, unpublished.




\end{references}
\end{document}